\documentclass[prd,twocolumn,showpacs,amsmath,amssymb,citeautoscript]{revtex4}

\def\cF{{c}_F}
\def\cS{{c}_\Sigma}
\def\cM{{c}_M}

\usepackage{graphicx,times}

\begin{document}

\title{Role of the $\sigma$-resonance in determining the convergence of chiral perturbation theory}
\author{D. J. Cecile and Shailesh Chandrasekharan}
\affiliation{ Department of Physics, Box 90305, Duke University,
Durham, North Carolina 27708.}

\begin{abstract}
The dimensionless parameter $\xi = M_\pi^2/(16 \pi^2 F_\pi^2)$, where $F_\pi$ is the pion decay constant and $M_\pi$ is the pion mass, is expected to control the convergence of chiral perturbation theory applicable to QCD. Here we demonstrate that a strongly coupled lattice gauge theory model with the same symmetries as two-flavor QCD but with a much lighter $\sigma$-resonance is different. Our model allows us to study efficiently the convergence of chiral perturbation theory as a function of $\xi$. We first confirm that the leading low energy constants appearing in the chiral Lagrangian are the same when calculated from the $p$-regime and the $\epsilon$-regime as expected. However, $\xi \lesssim 0.002$ is necessary before 1-loop chiral perturbation theory predicts the data within 1\%. For $\xi > 0.0035$ the data begin to deviate dramatically from 1-loop chiral perturbation theory predictions. We argue that this qualitative change is due to the presence of a light $\sigma$-resonance in our model. Our findings may be useful for lattice QCD studies.
\end{abstract}


\maketitle

Chiral perturbation theory has been successful in explaining a variety of experiments involving low energy pions \cite{Gasser:1983yg}. It is a low energy effective field theory that captures the chiral symmetry properties of QCD. The dynamical properties of QCD are encoded through a series of low energy constants. At the leading order there are two low energy constants: $F$ the pion decay constant and $\Sigma$ the chiral condensate, both evaluated in the chiral limit. One of the important topics of research today is to compute these and other higher order low energy constants from first principles using lattice QCD \cite{Necco:2007pr,Dimopoulos:2007qy,Bernard:2006zp,Cirigliano:2006hb}. Interestingly, the effects of a small quark mass $m$, which breaks the chiral symmetry explicitly, can also be taken into account and physical quantities can be expressed as a power series in a dimensionless parameter $\xi = M_\pi^2 /(16 \pi^2 F_\pi^2)$ where $M_\pi$ is the physical pion mass and $F_\pi$ is the physical pion decay constant. This series, which we refer to as the ``chiral expansion'', not only contains powers of $\xi$ but also powers of $\xi\log\xi$, $\xi^2\log\xi$ and so on. In QCD we can estimate $\xi \sim 0.015$ assuming $M_\pi \sim 140$ MeV and $F_\pi \sim 90$MeV. For later convenience we also define $\xi' = M^2/(16 \pi^2 F^2)$ where $M^2 = m \Sigma/F^2$ is the square of the pion mass to the leading order in the quark mass. It is easily verified that $\xi \approx \xi' + {\cal O}(\xi'^2)$, which means that to the first order in $\xi$ we can ignore the difference between $\xi$ and $\xi'$.

Given the smallness of $\xi$ it is not surprising that chiral perturbation theory works remarkably well in describing the physical world. An important question in the field is to find the range in $\xi$ where 1-loop perturbation theory will be valid up to a given error say $1$\% \cite{Sharpe:2006pu,Bernard:2002yk,Giusti:2007hk}. This will help lattice QCD calculations to extract reliably the low energy constants. Current lattice calculations typically use 2-loop chiral perturbation theory in the region $0.02 < \xi < 0.1$ to fit the data in order to extract the low energy constants of QCD \cite{Matsufuru:2007uc,Boyle:2007fn,Urbach:2007rt,Kuramashi:2007gs}. Is this reasonable? What is the physics that controls the convergence properties of the chiral expansion? Answers to such questions are crucial for future progress. Typically one believes that it is the $\rho$ meson resonance that puts the limit on pion masses where chiral perturbation theory will be valid. In order to avoid physically important singularities in $\pi-\pi$ scattering, it is reasonable to expect $M_\pi < M_\rho/2$ is necessary for chiral perturbation theory to be reliable. Experts believe that perhaps one needs at least $M_\pi < M_\rho/3$ \cite{Bernard:2002yk}. 

In principle, there is another resonance that can limit the convergence of the chiral expansion. This is the so called $\sigma$-resonance and arises in $\pi-\pi$ scattering in a channel with vacuum quantum numbers. Recently, the properties of this resonance in the physical world were estimated using experimental input, dispersion theory and chiral perturbation theory. It was estimated that $M_\sigma \simeq 440$MeV and $\Gamma_\sigma \simeq 544$MeV \cite{Caprini:2005zr}. This makes it a very broad and perhaps not so interesting resonance in the context of the convergence of chiral perturbation theory. On the other hand, in lattice QCD, as the pion masses increase, this resonance could become sharper and $\sigma$ could become a stable physical particle, a bound state of two pions. However, it could also remain an uninteresting resonance up to much higher pion masses. Recent studies find that the properties of the $\sigma$-resonance do depend strongly  on the quark mass \cite{Pelaez:2007,Hanhart:2008mx}. It is interesting to ask if this dependence can affect the chiral expansion. Although this is a difficult question to answer in QCD, it may be possible to explore it with simpler models. Here we show that the $\sigma$-resonance can in principle affect the chiral expansion. In particular we demonstrate that a light and weakly interacting $\sigma$ can induce an early break down of chiral perturbation theory.

It is easy to argue that a light $\sigma$-resonance can indeed trigger the breakdown of chiral perturbation theory. Consider a non-linear sigma model which contains a coupling $T$ that can be tuned such that for $T<T_c$ it is in a phase where the global symmetry is spontaneously broken and for $T>T_c$ it is in a symmetric phase. Chiral perturbation theory must be useful in describing the low energy properties of the theory in the broken phase, but not in the symmetric phase. This means, as $T$ is tuned towards $T_c$ in the broken phase, chiral perturbation theory must become poorly convergent. Close to $T_c$, if the phase transition is second order, the linear sigma model becomes a good description of the physics and in that model, as we will see later, the breakdown of chiral perturbation theory can be traced to the fact that $M_\sigma/F_\pi$ becomes small. Note that at the critical point the sigma and the pions become degenerate and chiral symmetry is completely restored. Although the scenario that a light $\sigma$-resonance affects the convergence of chiral perturbation theory is perhaps known to the experts, we do not know of any previous work which demonstrates this explicitly. This is the main motivation for our current work. Here we study a QCD-like lattice field theory model which has the same symmetries as two-flavor QCD. Hence $SU(2)\times SU(2)$ chiral perturbation theory is applicable. Our model also contains a parameter equivalent to the coupling $T$ of the non-linear sigma model discussed above. We tune this coupling to be close to the critical point and hence know that our model contains a light sigma resonance although we do not know its exact properties a priori. We then find evidence that indeed chiral perturbation theory breaks down when $M_\pi > M_\sigma/3$ is roughly satisfied.

Our model involves two flavors of staggered fermions interacting strongly with abelian gauge fields. We recently developed an efficient cluster algorithm for this model and studied it in the $\epsilon$-regime \cite{Cecile:2007dv}. Here we will focus on the $p$-regime. The action of the model is given by 
\begin{eqnarray}
S &=& - \sum_{x}\sum_{\mu=1}^{5}
\eta_{\mu,x}\bigg[\mathrm{e}^{i\phi_{\mu,x}}{\overline\psi}_x
{\psi}_{x+\hat\mu}
-\mathrm{e}^{-i\phi_{\mu,x}}{\overline\psi}_{x+\hat\mu}{\psi}_x\bigg]
\nonumber
\\
&& - \sum_x \bigg[m{\overline\psi}_x{\psi}_x
+\frac{\tilde c}{2}\bigg({\overline\psi}_x{\psi}_x \bigg)^2\bigg],
\label{eq1}
\end{eqnarray}
where $x$ denotes a lattice site on a $4+1$ dimensional hyper-cubic lattice $L_t \times L^4$.  Here $L^4$ is the usual Euclidean space-time box while $L_t$ represents a fictitious temperature direction whose role will be discussed below. The two component Grassmann fields, $\overline\psi_x$ and $\psi_x$, represent the two quark $(u,d)$ flavors of mass $m$, and $\phi_{\mu,x}$ is the compact $U(1)$ gauge field through which the quarks interact. Here $\mu = 1,2,..,5$  runs over the $4+1$ directions. The $\mu=1$ direction will denote the fictitious temperature direction, while the remaining directions represent Euclidean space-time. The usual staggered fermion phase factors $\eta_{\mu,x}$ obey the relations: $\eta_{1,x}^2 = T$ and $\eta_{i,x}^2 = 1$ for $i=2,3,4,5$. The parameter $T$ controls the fictitious temperature. The four fermion coupling $\tilde c$ sets the strength of the anomaly. As explained in \cite{Cecile:2007dv}, the above model has the same symmetries as $N_f=2$ QCD, i.e..,  when $m = 0$, the action exhibits a global $SU_L(2)\times SU_R(2)$ symmetry, which is explicitly broken down to $SU_V(2)$ when $m\neq 0$. In this work we fix $L_t=2$ and $\tilde{c} = 0.3$. For these parameters the temperature $T$ can be tuned so that the model is in a spontaneously broken phase for $T<T_c$ or in the symmetric phase for $T>T_c$, where $T_c=1.73779(4)$ was determined in the earlier work \cite{Cecile:2007dv}. Since the phase transition is second order, close to $T_c$ the pion decay constant in the chiral limit $F$ is small in lattice units. This reduces the lattice artifacts in our model. Further, tuning $T$ close to $T_c$ also makes the $\sigma$-resonance light as discussed above. For these reasons, we chose to fix $T=1.7$ in this work.

We focus on three observables: The vector current susceptibility $Y_v$, the chiral current susceptibility $Y_c$ and the chiral condensate susceptibility $\chi_\sigma$. The current susceptibilities are defined as
\begin{equation}
\label{yv}
Y_{v,c} = \frac{1}{d L^d}\bigg\langle 
\sum_{\mu=1}^d\bigg(\sum_x J_{\mu}^{v,c}(x)\bigg)^2\bigg\rangle
\end{equation}
where $J_\mu^{v}(x)$ and $J_\mu^c(x)$ denote one of the components of the vector and the chiral current respectively. The condensate susceptibility is defined as
\begin{equation}
\chi_\sigma = \frac{1}{L^d} \sum_{x,y} 
\langle \overline{\psi}_x\psi_x \ \overline{\psi}_y\psi_y\rangle 
\end{equation}
For a detailed discussion of our algorithm and observables, we refer the reader to \cite{Cecile:2007dv}.

The behavior of these observables for large $L$ and small $m$ is governed by chiral perturbation theory which is described by the Euclidean chiral Lagrangian density
\begin{equation}
{\cal L} = 
\frac{F^2}{4}
\mathrm{tr}\Big(\partial_\mu U^\dagger \partial_\mu U\Big)
- \frac{m \Sigma}{4} tr \Big(U + U^\dagger \Big)
\end{equation}
where $F$ is the chiral pion decay constant, $\Sigma$ is the chiral condensate and $U\in SU(2)$ is the pion field. Using this Lagrangian, the finite size scaling formulas for many quantities have been found in in the literature \cite{Hasenfratz:1989pk,Hansen:1990yg,Colangelo:2003hf,Colangelo:2005gd,Colangelo:2006mp}. The predictions for $Y_c$, $Y_v$ and $\chi_\sigma$ in the $p$-regime can be found in \cite{Hansen:1990yg,Colangelo:2006mp}:
\begin{subequations}
\label{fss}
\begin{eqnarray}
Y_c &=& (F_\pi)^2 \Bigg[1 - 2 \tilde{g_1}(L M_\pi)\xi + {\cal O}(\xi^2)\Bigg]
\\ \nonumber \\ 
Y_v &=&  (F_\pi)^2 \Bigg[ - 2 L \frac{\partial \tilde{g_1}(L M_\pi)}{\partial L} \xi + {\cal O}(\xi^2)\Bigg]
\\ \nonumber \\
\chi_\sigma &=& (\langle{\overline q}q\rangle)^2 L^4
\bigg[1 - 3 \tilde{g_1}(L M_\pi) \xi + {\cal O}(\xi^2)\bigg]  
\end{eqnarray}
\end{subequations}
where $M_\pi$ is the pion mass, $F_\pi$ is the pion decay constant and $\langle{\overline q}q\rangle$ is the chiral condensate at a given quark mass $m$. The function $\tilde{g_1}$ arises due to pions constrained to be inside a periodic box and is given by
\begin{equation}
\label{gfn}
\tilde{g}_1(\lambda) = \sum_{n_1,n_2,n_3,n_4\neq 0}^{\infty}
\frac{4}{\lambda \sqrt n} K_1(\lambda \sqrt n)
\end{equation}
where $K_1$ is a Bessel function of the second kind and $n = n_1^2 + n_2^2 + n_3^2 +n_4^2$.

\begin{figure}[t]
\begin{center}
\vbox{
\includegraphics[width=0.48\textwidth]{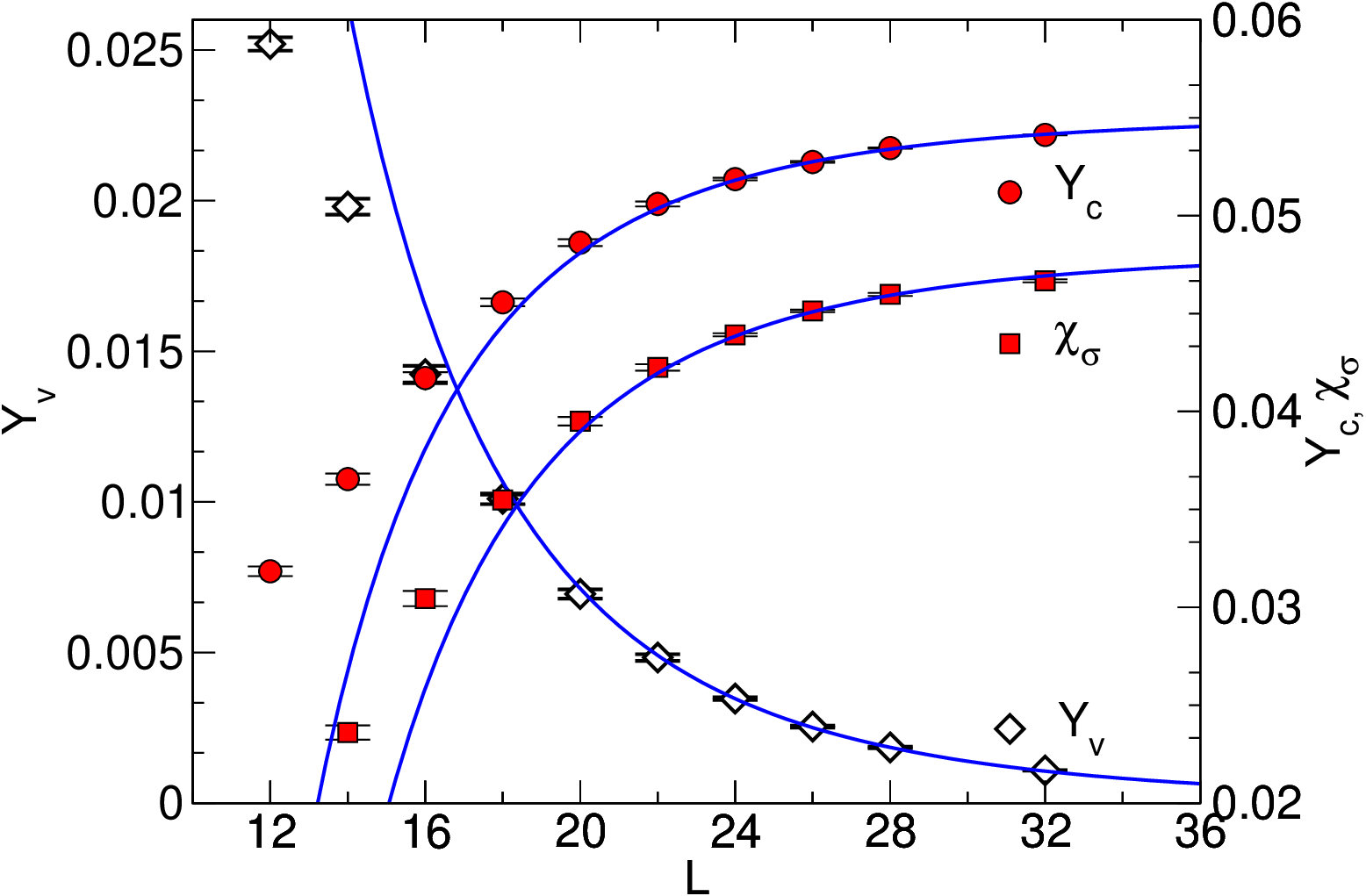}
\vskip0.4in
\includegraphics[width=0.48\textwidth]{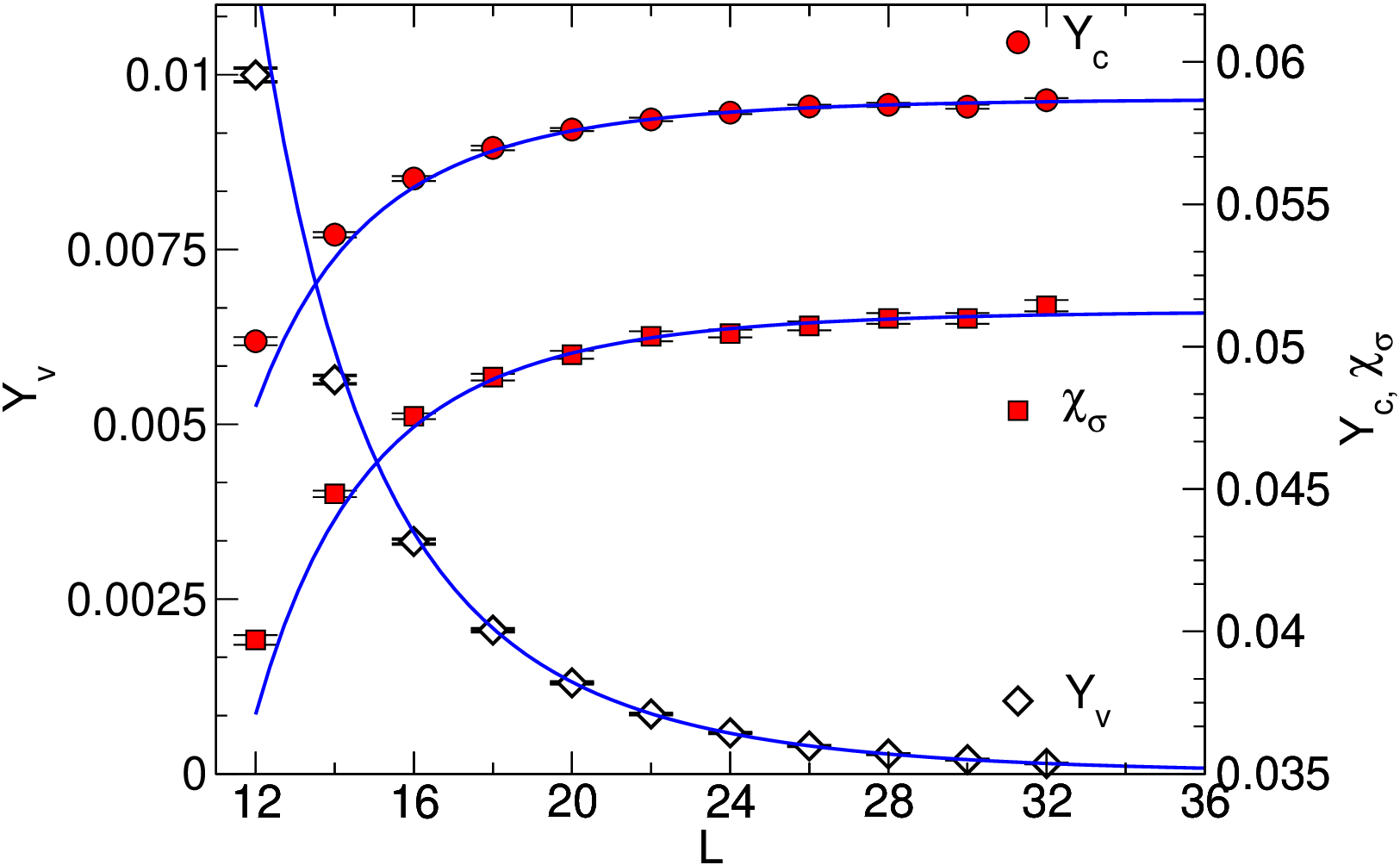}
}
\end{center}
\caption{\label{fig1} Finite size scaling of $Y_v$,$Y_c$ and $\chi_\sigma$ at $m=0.0002$ (top) and $m=0.001$ (bottom). The solid lines are fits of the data to the expected form from chiral perturbation theory.}
\end{figure}

We have varied the quark mass in the interval $0.0002 \leq m \leq 0.01$ for lattices in the range $12 \leq L \leq 32$. Our data fits well to the above predictions of chiral perturbation theory for $0.0002 < m \leq 0.0035$. The detailed results are summarized in Table \ref{tab1}.
\begin{table}[h]
\begin{center}
\begin{tabular}{c| c c c |c c} \hline\hline
\em m &\em $\langle \overline q q \rangle$ &\em $F_\pi$ &\em $M_\pi$ &\em
$\chi^2$ &\em Fit range
\\\hline
0.0002 &0.4392(2) &0.2348(1) &0.0400(2) &2.5 &$24\leq L \leq 32$ \\
0.0005 &0.4441(2) &0.2377(1) &0.0627(2) &1.1 &$24\leq L \leq 32$ \\
0.0008 &0.4499(2) &0.2406(1) &0.0789(1) &0.9 &$22\leq L \leq 32$ \\
0.0010 &0.4528(2) &0.2423(1) &0.0878(1) &0.8 &$18\leq L \leq 32$ \\
0.0015 &0.4606(2) &0.2467(1) &0.1070(2) &1.3 &$18\leq L \leq 32$ \\
0.0020 &0.4678(2) &0.2501(1) &0.1220(2) &1.8 &$20\leq L\leq 32$ \\
0.0025 &0.4740(2) &0.2538(1) &0.1356(2) &1.6 &$16\leq L\leq 32$ \\
0.0035 &0.4867(2) &0.2606(1) &0.1584(2) &0.9 &$16\leq L\leq 32$ \\
\hline\hline
\end{tabular}
\end{center}
\caption{Results from fitting $Y_v$,$Y_c$, and $\chi_\sigma$ as a function of $L$ to the finite-size one-loop chiral perturbation theory. The $\chi^2$ quoted is per degree of freedom \label{tab1}}
\end{table}
Thus, we are able to extract $F_\pi$, $M_\pi$ and $\langle {\overline q}q\rangle$ as functions of the quark mass. As an illustration, we show the data at $m=0.0002$ and $m=0.001$ in Fig.~\ref{fig1}. Note that the fit is not as reliable at the lowest mass ($m=0.0002$) as compared to higher masses. It is possible that our lattices are not sufficiently large at this tiny quark mass to allow us to fit to 1-loop results.

\begin{figure}[t]
\begin{center}
\vbox{
\includegraphics[width=0.48\textwidth]{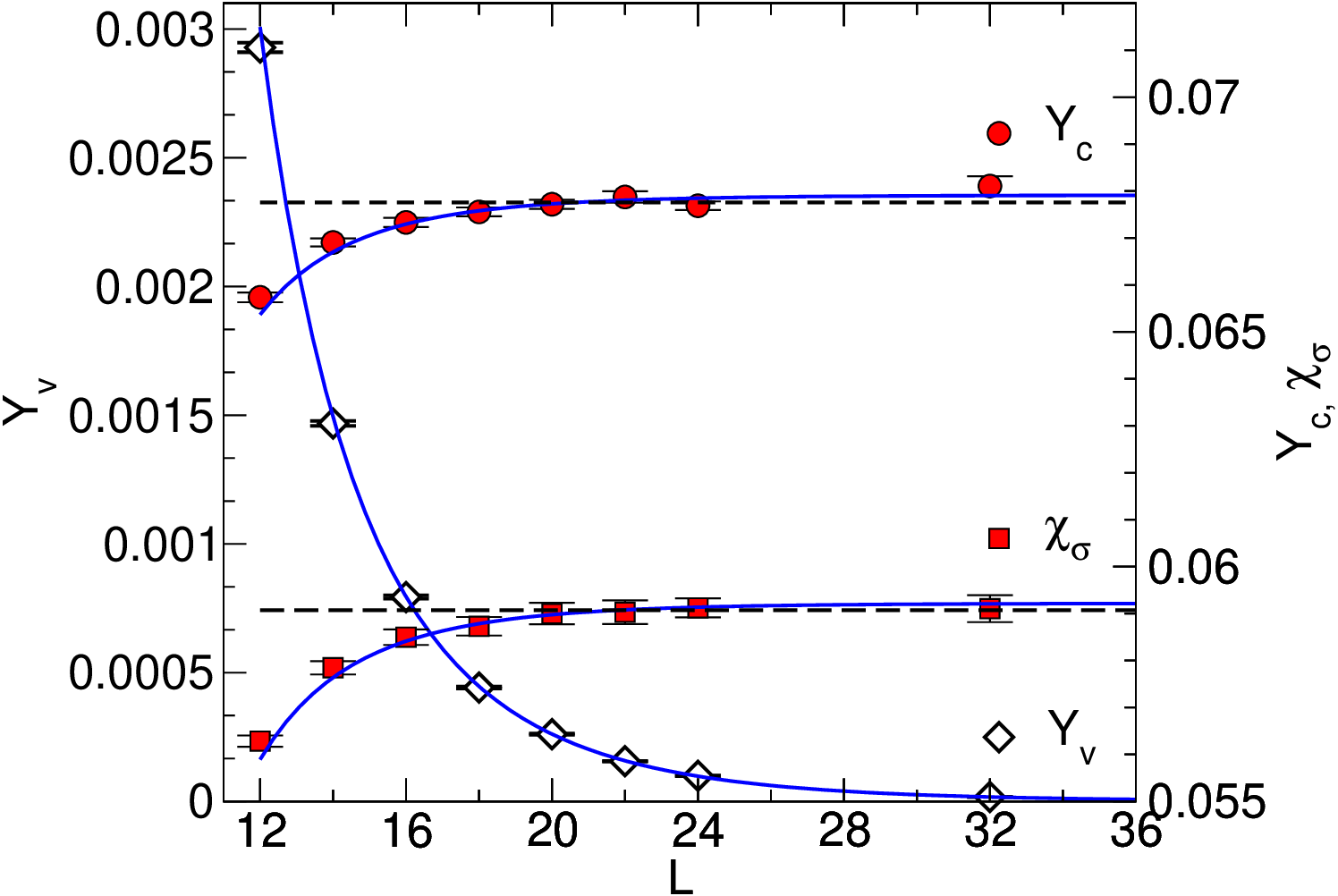}
\vskip0.4in
\includegraphics[width=0.48\textwidth]{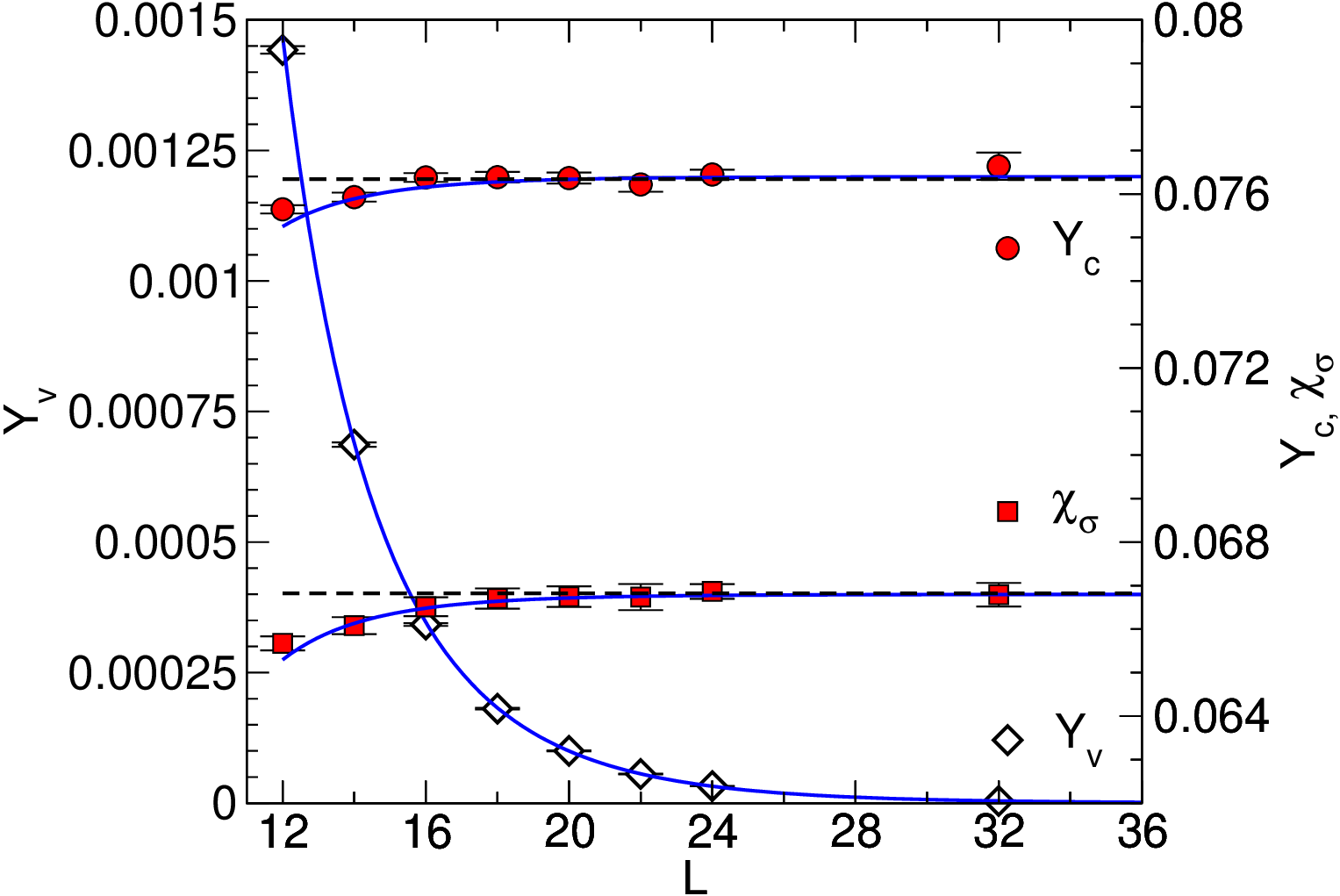}
}
\end{center}
\caption{\label{fig2} Finite size scaling of $Y_v$,$Y_c$ and $\chi_\sigma$ at $m=0.0035$ (top) and $m=0.0065$ (bottom). The solid lines are fits of the data to the expected finite size scaling form from chiral perturbation theory while dashed lines are fits to a constant.}
\end{figure}

At $m \geq 0.002$ the fits converge only if we exclude almost all the curvature in $Y_c$ and $\chi_\sigma$. In particular, we are not sensitive to the $\tilde{g}_1(\lambda)$ function for these two observables and the data fit well even to a constant as shown in Table \ref{tab2}.
\begin{table}[h]
\begin{center}
\begin{tabular}{c| c c | c c | c c} \hline\hline
\em m &\em $\langle \overline q q \rangle$ &\em $\chi^2$ &\em
$F_\pi$ &\em $\chi^2$
&\em $M_\pi$ &\em $\chi^2$\\
\hline
0.0020 &0.4668(3) &1.2 &0.2498(1) &0.1 &0.1226(2) &0.6  \\
0.0025 &0.4728(3) &0.7 &0.2536(2) &0.9 &0.1356(2) &1.6  \\
0.0035 &0.4861(3) &0.1 &0.2603(1) &1.5 &0.1584(2) &1.7  \\
0.0050 &0.5024(3) &0.2 &0.2690(2) &1.1 &0.1860(3) &0.7  \\
0.0065 &0.5170(3) &0.1 &0.2764(2) &0.7 &0.2083(4) &0.5  \\
0.0075 &0.5247(3) &0.2 &0.2807(2) &1.6 &0.2219(4) &0.9  \\
0.0100 &0.5433(2) &0.7 &0.2912(2) &0.1 &0.2521(5) &1.8  \\
\hline\hline
\end{tabular}
\end{center}
\caption{Results from fitting $Y_c$ and $\chi_\sigma$ to a constant while $Y_v$ is fit to one-loop chiral perturbation theory.\label{tab2}}
\end{table}
We illustrate this issue by plotting the data and the fits at $m=0.0035$ and $0.0065$ in Fig.~\ref{fig2}. Comparing the results from the two different fits we see that the error bars for $\langle \overline q q \rangle$ are underestimated by a factor of two or three at the higher masses. We find that $M_\pi$ can be calculated very accurately by a one-parameter fit of $Y_v$ which may be a useful observation for lattice QCD calculations. Interestingly, $Y_v$ continues to fit well to the one-loop formula even at higher masses, but $n$ (in Eq.~\ref{gfn}) could be restricted to small values (typically less than 3).

\begin{table}[b]
\begin{center}
\begin{tabular}{c  c   c   c   c | c } \hline\hline
$\Sigma$ & $F$ &\em $\cS$ & $\cF$ & $\cM$ &\em $\chi^2$
\\\hline
0.4354(3) &0.2329(2) &11.9(3) &19.3(5) &39(3) &1.1    \\
0.4351(5) &0.2331(4) &12.3(5) &18.9(9) &37(3) &1.6    \\
\hline\hline
\end{tabular}
\end{center}
\caption{Results from a combined fit of the data in Table \ref{tab1} to Eqs.~\ref{infvol}. The first row uses data in the range $0.0002 \leq m \leq 0.001$ while the second row excludes $m=0.0002$ from the fit.\label{tab3}}
\end{table}

The quark mass dependence of $F_\pi$, $\langle \overline{q}q\rangle$ and $M_\pi$ have been computed up to 1-loop in \cite{Hansen:1990yg,Hasenfratz:1989pk}:
\begin{subequations}
\label{infvol}
\begin{eqnarray}
F_\pi &=& F\bigg[1 -\xi'\log\xi' + 2\xi' \cF\bigg] \\ 
\nonumber \\
\langle{\overline q}q\rangle &=& 
\Sigma\bigg[1 - \frac{3}{2}\xi' \log\xi' + 3\xi' \cS \bigg] 
\\ \nonumber \\
M^2_\pi &=& M^2 
\bigg[1 +\frac{1}{2}\xi'\log\xi' - \xi' \cM\bigg].
\end{eqnarray}
\end{subequations}
where $\cF,\cS$ and $\cM$ are higher order low energy constants and are usually defined in the literature as $c_i = \log(\Lambda_i/4\pi F)$. We have performed a combined fit of all the values of $F_\pi$,$\langle{\overline q}q\rangle$ and $M_\pi$ quoted in Table \ref{tab1} in the region $0.0002 \leq m \leq 0.001$ to the above three relations. The result is tabulated in the first row of the Table \ref{tab3}. We note that the values of $F$ and $\Sigma$ agree nicely with $F=0.2327(1)$ and $\Sigma=0.4346(2)$ computed earlier at $m=0$ \cite{Cecile:2007dv}. Further, in the $\epsilon$-regime we find $\cM+4\cS = 80(6)$, while in the $p$-regime (from Table \ref{tab3}) we see that this number is $87(4)$. Thus, we confirm that the $p$-regime and the $\epsilon$-regime are described by the same low energy constants as expected. {\em This is the first important result of our work}.

\begin{figure}[t]
\begin{center}
\vbox{
\includegraphics[width=0.45\textwidth]{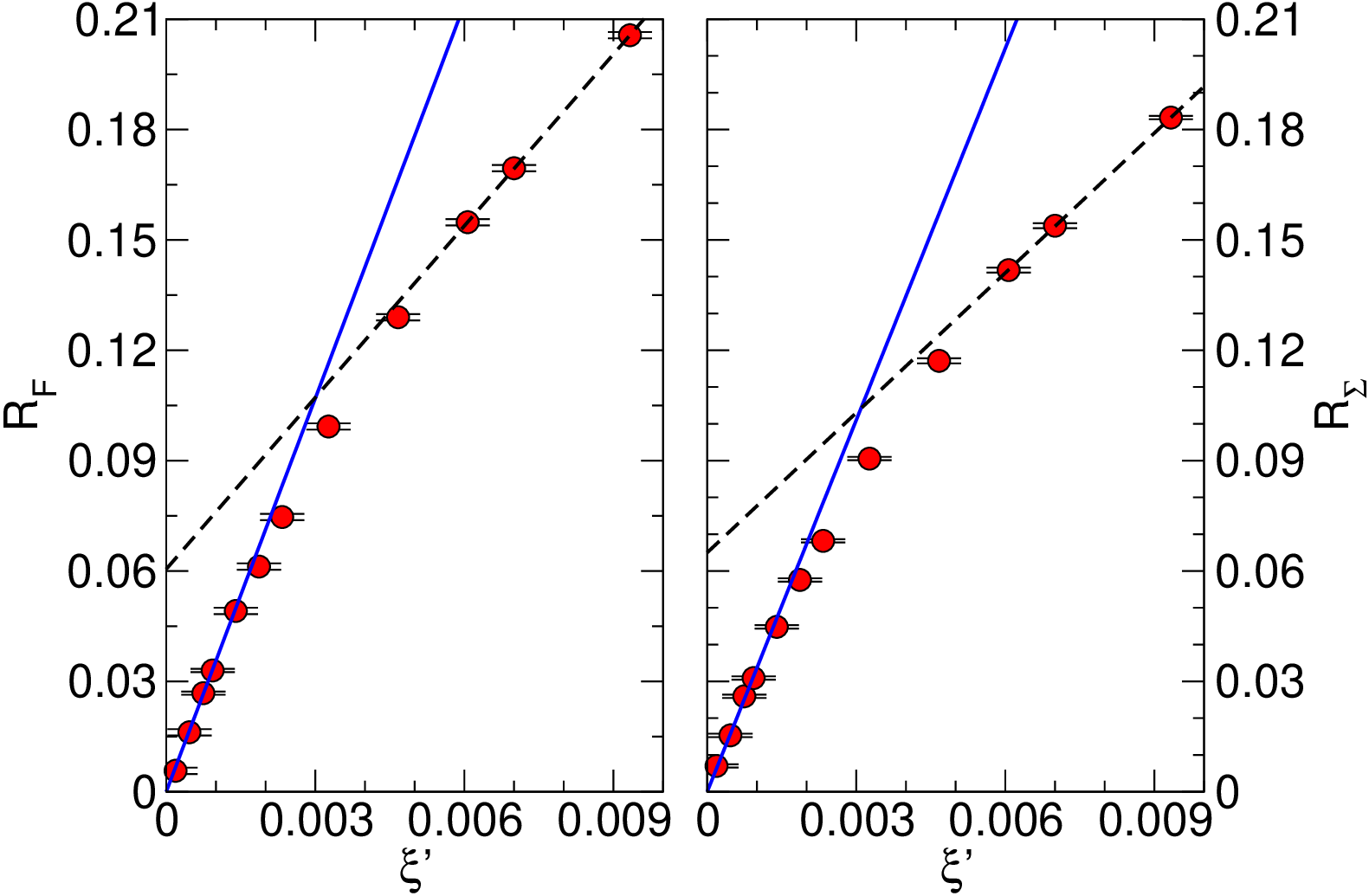}
\vskip0.4in
\includegraphics[width=0.45\textwidth]{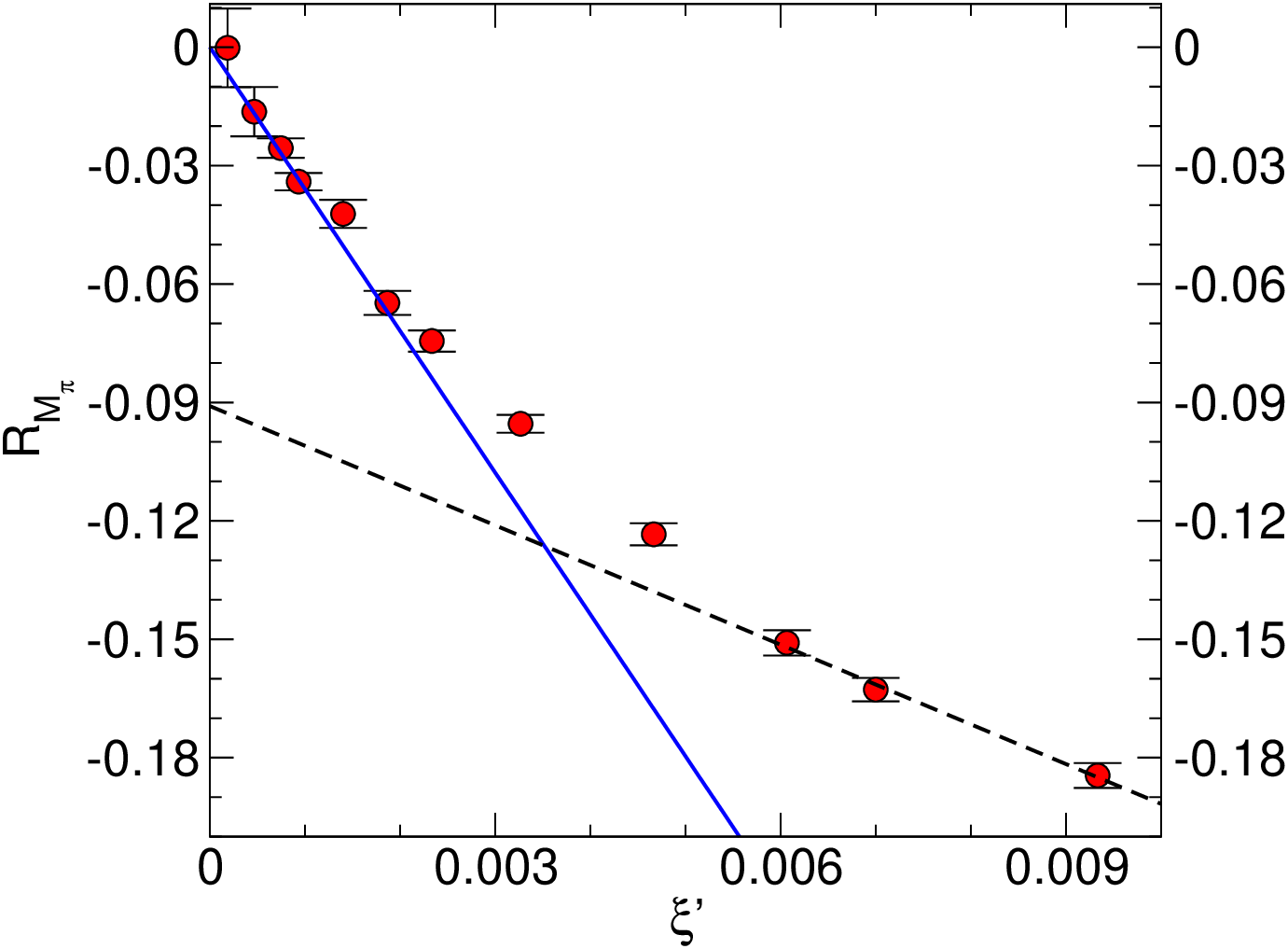}
}
\end{center}
\caption{\label{fig3} Rescaled and subtracted quantities $R_{F_\pi}$, $R_{\langle{\overline q}q\rangle}$ and $R_{M^2_\pi}$ defined in Eqs.~(\ref{reqs}). The solid lines are plots of the fits discussed in the text. The dashed lines show the linear region for larger values of $\xi'$. The ``knee'' is estimated roughly as the point where the two lines cross.}
\end{figure}

In order to isolate the region where one-loop corrections are a good description of the data we define the following rescaled and subtracted quantities:
\begin{subequations}
\label{reqs}
\begin{eqnarray}
R_{F} &\equiv& F_\pi/F - 1 + \xi'\log\xi', \\
\nonumber \\
R_{\Sigma} &\equiv& 
\langle{\overline q}q\rangle/\Sigma - 1 + 3\xi'\log\xi'/2,\\
\nonumber \\
R_{M} &\equiv& M_\pi^2/M^2 - 1 - \xi'\log(\xi')/2. 
\end{eqnarray}
\end{subequations}
We use chiral values $F=0.2329$ and $\Sigma =0.4354$ to compute the $R$'s and $\xi'$. By definition, the $R$'s must be linear in $\xi'$ in the region where one-loop results are valid. In Fig.~\ref{fig3} we plot the $R$'s as a function of $\xi'$. Assuming errors of $1\%$ or less can be tolerated, Fig.~\ref{fig3} shows that the linear region of 1-loop chiral perturbation theory occurs roughly when $\xi' \lesssim 0.002$. Interestingly, there is also an approximately linear region for $\xi' \gtrsim 0.006$ but with a completely different slope. This is shown as the dashed line in Fig.~\ref{fig3}. This behavior suggests that chiral perturbation theory begins to break down. We will argue below that the $\sigma$-resonance is responsible for this break down. Note that $\xi' \approx 0.0035$ is the rough location of the ``knee'' that separates the low $\xi'$ and high $\xi'$ regions.

The unnaturally large values of $\cF$, $\cS$ and $\cM$ are clearly responsible for the break down of the chiral expansion at very small values of $\xi'$. What is the physics behind these large values? It has been argued in the context of the $O(4)$ linear sigma model, that the physics in the sigma channel is directly related to these terms. In particular, perturbative calculations show that \cite{Gockeler:1992zj,Gockeler:1991sj,Hasenfratz:1990fu}:
\begin{subequations}
\label{sigma}
\begin{eqnarray}
\cS &=& \log(M_R/4\pi F) - \frac{7}{6} + \frac{8\pi^2}{3 g_R} 
\\
\cM &=& \log(M_R/4\pi F) - \frac{7}{3} + \frac{8\pi^2}{g_R} 
\end{eqnarray}
\end{subequations}
where 
\begin{equation}
M^2_\sigma = M^2_R \bigg[1 + \frac{g_R}{16\pi^2} 
\big(3\pi\sqrt 3 -13 \big)\bigg]
\end{equation}
Here $M_\sigma$ is that mass of the $\sigma$ particle and $g_R$ is the corresponding renormalized coupling, $g_R = M_R^2/2 F^2$. We believe that in our model the above relations must be valid at least as a good approximation because we are close to the critical point where $g_R$ is expected to be small and the perturbative $O(4)$ linear sigma model is a good description of the low energy physics. Indeed, using $\cS = 12$ we find that $M_\sigma/F \sim 2$ while using $\cM = 39$ we again find that $M_\sigma/F \sim 2$. The fact that these two agree with each other is a clear vindication of our belief. Assuming $M_\sigma/F \sim 2$ and setting the scale of our lattice with $F = 90$MeV we estimate $M_\sigma \sim 180$MeV in our model. At $\xi' \sim 0.0035$ we find that $M_\pi \sim 60$MeV. Hence, we conclude that when $M_\pi > M_\sigma/3$ chiral perturbation theory begins to break down and the physics is better described by the linear sigma model. {\em This is the second important result of our work.}

Can we learn something about QCD from our work? Although there are many important differences between our model and QCD, the main difference is that we have tuned our model so that it contains a light and most likely narrow $\sigma$-resonance. In QCD the $\sigma$ is expected to be heavier and broader. We think this is the difference why our low energy constants turned out to be much larger than QCD, which in turn affected the convergence of chiral perturbation theory. Clearly, at the minimum we have learned that the properties of the $\sigma$-resonance do affect the low energy constants of the chiral expansion. These properties do change with the quark mass while, by definition, the low energy constants are independent of the quark mass. This suggests that chiral perturbation theory may become reliable only in the region of the quark mass where the properties of the $\sigma$-resonance do not change much. This is the most important lesson of relevance to QCD from our work. The quark mass dependence of the $\sigma$ and the $\rho$ was recently studied in \cite{Hanhart:2008mx}. In particular it was found that the coupling of $\sigma$ to the pions changes significantly with $M_\pi$. A rough estimate suggests that $M_\pi \lesssim 250$MeV is necessary for the properties of the $\sigma$-resonance to become stable. It would be interesting if this is also the region where chiral perturbation theory becomes a reliable tool.

In summary, here we have studied a model with the same symmetries as $N_f=2$ QCD from first principles and have shown that chiral perturbation theory is a reliable tool only for small quark masses. In particular we learned that the $\sigma$-resonance can be important in determining the convergence of the chiral expansion. Studying the quark mass dependence of the $\sigma$-resonance should be useful and can in principle be done using lattice QCD and may shed light on the region of quark masses where 1-loop chiral perturbation theory is valid in QCD up to a given error.

We thank G. Colangelo for useful discussions about the $\sigma$-resonance. We also thank C.Bernard, S.~D\"{u}rr, C.~Haefeli, F.-J.~Jiang, H.~Leutwyler, T.~Mehen, K.~Orginos, B.~Tiburzi and U.-J.~Wiese for helpful comments. This work was supported in part by the Department of Energy grant DE-FG02-05ER41368.

\bibliography{chpt}

\begin{thebibliography}{24}
\expandafter\ifx\csname natexlab\endcsname\relax\def\natexlab#1{#1}\fi
\expandafter\ifx\csname bibnamefont\endcsname\relax
  \def\bibnamefont#1{#1}\fi
\expandafter\ifx\csname bibfnamefont\endcsname\relax
  \def\bibfnamefont#1{#1}\fi
\expandafter\ifx\csname citenamefont\endcsname\relax
  \def\citenamefont#1{#1}\fi
\expandafter\ifx\csname url\endcsname\relax
  \def\url#1{\texttt{#1}}\fi
\expandafter\ifx\csname urlprefix\endcsname\relax\def\urlprefix{URL }\fi
\providecommand{\bibinfo}[2]{#2}
\providecommand{\eprint}[2][]{\url{#2}}

\bibitem[{\citenamefont{Gasser and Leutwyler}(1984)}]{Gasser:1983yg}
\bibinfo{author}{\bibfnamefont{J.}~\bibnamefont{Gasser}} \bibnamefont{and}
  \bibinfo{author}{\bibfnamefont{H.}~\bibnamefont{Leutwyler}},
  \bibinfo{journal}{Ann. Phys.} \textbf{\bibinfo{volume}{158}},
  \bibinfo{pages}{142} (\bibinfo{year}{1984}).

\bibitem[{\citenamefont{Necco}(2007)}]{Necco:2007pr}
\bibinfo{author}{\bibfnamefont{S.}~\bibnamefont{Necco}} (\bibinfo{year}{2007}),
  \eprint{arXiv:0710.2444 [hep-lat]}.

\bibitem[{\citenamefont{Dimopoulos et~al.}(2007)\citenamefont{Dimopoulos,
  Frezzotti, Herdoiza, Urbach, and Wenger}}]{Dimopoulos:2007qy}
\bibinfo{author}{\bibfnamefont{P.}~\bibnamefont{Dimopoulos}},
  \bibinfo{author}{\bibfnamefont{R.}~\bibnamefont{Frezzotti}},
  \bibinfo{author}{\bibfnamefont{G.}~\bibnamefont{Herdoiza}},
  \bibinfo{author}{\bibfnamefont{C.}~\bibnamefont{Urbach}}, \bibnamefont{and}
  \bibinfo{author}{\bibfnamefont{U.}~\bibnamefont{Wenger}}
  (\bibinfo{collaboration}{ETM}) (\bibinfo{year}{2007}),
  \eprint{arXiv:0710.2498 [hep-lat]}.

\bibitem[{\citenamefont{Bernard et~al.}(2006)}]{Bernard:2006zp}
\bibinfo{author}{\bibfnamefont{C.}~\bibnamefont{Bernard}} \bibnamefont{et~al.}
  (\bibinfo{year}{2006}), \eprint{hep-lat/0611024}.

\bibitem[{\citenamefont{Cirigliano et~al.}(2006)}]{Cirigliano:2006hb}
\bibinfo{author}{\bibfnamefont{V.}~\bibnamefont{Cirigliano}}
  \bibnamefont{et~al.}, \bibinfo{journal}{Nucl. Phys.}
  \textbf{\bibinfo{volume}{B753}}, \bibinfo{pages}{139} (\bibinfo{year}{2006}),
  \eprint{hep-ph/0603205}.

\bibitem[{\citenamefont{Sharpe}(2006)}]{Sharpe:2006pu}
\bibinfo{author}{\bibfnamefont{S.~R.} \bibnamefont{Sharpe}}
  (\bibinfo{year}{2006}), \eprint{hep-lat/0607016}.

\bibitem[{\citenamefont{Bernard et~al.}(2003)}]{Bernard:2002yk}
\bibinfo{author}{\bibfnamefont{C.}~\bibnamefont{Bernard}} \bibnamefont{et~al.},
  \bibinfo{journal}{Nucl. Phys. Proc. Suppl.} \textbf{\bibinfo{volume}{119}},
  \bibinfo{pages}{170} (\bibinfo{year}{2003}), \eprint{hep-lat/0209086}.

\bibitem[{\citenamefont{Giusti}(2007)}]{Giusti:2007hk}
\bibinfo{author}{\bibfnamefont{L.}~\bibnamefont{Giusti}},
  \bibinfo{journal}{PoS.} \textbf{\bibinfo{volume}{LAT2006}}
  (\bibinfo{year}{2007}), \eprint{hep-lat/0702014}.

\bibitem[{\citenamefont{Matsufuru}(2007)}]{Matsufuru:2007uc}
\bibinfo{author}{\bibfnamefont{H.}~\bibnamefont{Matsufuru}}
  (\bibinfo{collaboration}{JLQCD}) (\bibinfo{year}{2007}),
  \eprint{arXiv:0710.4225 [hep-lat]}.

\bibitem[{\citenamefont{Boyle}(2007)}]{Boyle:2007fn}
\bibinfo{author}{\bibfnamefont{P.}~\bibnamefont{Boyle}}
  (\bibinfo{collaboration}{RBC}) (\bibinfo{year}{2007}),
  \eprint{arXiv:0710.5880 [hep-lat]}.

\bibitem[{\citenamefont{Urbach}(2007)}]{Urbach:2007rt}
\bibinfo{author}{\bibfnamefont{C.}~\bibnamefont{Urbach}}
  (\bibinfo{year}{2007}), \eprint{arXiv:0710.1517 [hep-lat]}.

\bibitem[{\citenamefont{Kuramashi}(2007)}]{Kuramashi:2007gs}
\bibinfo{author}{\bibfnamefont{Y.}~\bibnamefont{Kuramashi}}
  (\bibinfo{year}{2007}), \eprint{arXiv:0711.3938 [hep-lat]}.

\bibitem[{\citenamefont{Caprini et~al.}(2006)\citenamefont{Caprini, Colangelo,
  and Leutwyler}}]{Caprini:2005zr}
\bibinfo{author}{\bibfnamefont{I.}~\bibnamefont{Caprini}},
  \bibinfo{author}{\bibfnamefont{G.}~\bibnamefont{Colangelo}},
  \bibnamefont{and}
  \bibinfo{author}{\bibfnamefont{H.}~\bibnamefont{Leutwyler}},
  \bibinfo{journal}{Phys. Rev. Lett.} \textbf{\bibinfo{volume}{96}},
  \bibinfo{pages}{132001} (\bibinfo{year}{2006}), \eprint{hep-ph/0512364}.

\bibitem[{\citenamefont{Pelaez et~al.}(2007)\citenamefont{Pelaez, Hanhart, and
  Rios}}]{Pelaez:2007}
\bibinfo{author}{\bibfnamefont{J.}~\bibnamefont{Pelaez}},
  \bibinfo{author}{\bibfnamefont{C.}~\bibnamefont{Hanhart}}, \bibnamefont{and}
  \bibinfo{author}{\bibfnamefont{G.}~\bibnamefont{Rios}}
  (\bibinfo{year}{2007}), \eprint{0712.1734}.

\bibitem[{\citenamefont{Hanhart et~al.}(2008)\citenamefont{Hanhart, Pelaez, and
  Rios}}]{Hanhart:2008mx}
\bibinfo{author}{\bibfnamefont{C.}~\bibnamefont{Hanhart}},
  \bibinfo{author}{\bibfnamefont{J.~R.} \bibnamefont{Pelaez}},
  \bibnamefont{and} \bibinfo{author}{\bibfnamefont{G.}~\bibnamefont{Rios}}
  (\bibinfo{year}{2008}), \eprint{arXiv:0801.2871 [hep-ph]}.

\bibitem[{\citenamefont{Cecile and Chandrasekharan}(2008)}]{Cecile:2007dv}
\bibinfo{author}{\bibfnamefont{D.~J.} \bibnamefont{Cecile}} \bibnamefont{and}
  \bibinfo{author}{\bibfnamefont{S.}~\bibnamefont{Chandrasekharan}},
  \bibinfo{journal}{Phys. Rev.} \textbf{\bibinfo{volume}{D77}},
  \bibinfo{eid}{014506} (pages~\bibinfo{numpages}{10}) (\bibinfo{year}{2008}).

\bibitem[{\citenamefont{Hasenfratz and Leutwyler}(1990)}]{Hasenfratz:1989pk}
\bibinfo{author}{\bibfnamefont{P.}~\bibnamefont{Hasenfratz}} \bibnamefont{and}
  \bibinfo{author}{\bibfnamefont{H.}~\bibnamefont{Leutwyler}},
  \bibinfo{journal}{Nucl. Phys.} \textbf{\bibinfo{volume}{B343}},
  \bibinfo{pages}{241} (\bibinfo{year}{1990}).

\bibitem[{\citenamefont{Hansen and Leutwyler}(1991)}]{Hansen:1990yg}
\bibinfo{author}{\bibfnamefont{F.~C.} \bibnamefont{Hansen}} \bibnamefont{and}
  \bibinfo{author}{\bibfnamefont{H.}~\bibnamefont{Leutwyler}},
  \bibinfo{journal}{Nucl. Phys.} \textbf{\bibinfo{volume}{B350}},
  \bibinfo{pages}{201} (\bibinfo{year}{1991}).

\bibitem[{\citenamefont{Colangelo and Durr}(2004)}]{Colangelo:2003hf}
\bibinfo{author}{\bibfnamefont{G.}~\bibnamefont{Colangelo}} \bibnamefont{and}
  \bibinfo{author}{\bibfnamefont{S.}~\bibnamefont{Durr}},
  \bibinfo{journal}{Eur. Phys. J.} \textbf{\bibinfo{volume}{C33}},
  \bibinfo{pages}{543} (\bibinfo{year}{2004}), \eprint{hep-lat/0311023}.

\bibitem[{\citenamefont{Colangelo et~al.}(2005)\citenamefont{Colangelo, Durr,
  and Haefeli}}]{Colangelo:2005gd}
\bibinfo{author}{\bibfnamefont{G.}~\bibnamefont{Colangelo}},
  \bibinfo{author}{\bibfnamefont{S.}~\bibnamefont{Durr}}, \bibnamefont{and}
  \bibinfo{author}{\bibfnamefont{C.}~\bibnamefont{Haefeli}},
  \bibinfo{journal}{Nucl. Phys.} \textbf{\bibinfo{volume}{B721}},
  \bibinfo{pages}{136} (\bibinfo{year}{2005}), \eprint{hep-lat/0503014}.

\bibitem[{\citenamefont{Colangelo and Haefeli}(2006)}]{Colangelo:2006mp}
\bibinfo{author}{\bibfnamefont{G.}~\bibnamefont{Colangelo}} \bibnamefont{and}
  \bibinfo{author}{\bibfnamefont{C.}~\bibnamefont{Haefeli}},
  \bibinfo{journal}{Nucl. Phys.} \textbf{\bibinfo{volume}{B744}},
  \bibinfo{pages}{14} (\bibinfo{year}{2006}), \eprint{hep-lat/0602017}.

\bibitem[{\citenamefont{Gockeler et~al.}(1993)\citenamefont{Gockeler, Kastrup,
  Neuhaus, and Zimmermann}}]{Gockeler:1992zj}
\bibinfo{author}{\bibfnamefont{M.}~\bibnamefont{Gockeler}},
  \bibinfo{author}{\bibfnamefont{H.~A.} \bibnamefont{Kastrup}},
  \bibinfo{author}{\bibfnamefont{T.}~\bibnamefont{Neuhaus}}, \bibnamefont{and}
  \bibinfo{author}{\bibfnamefont{F.}~\bibnamefont{Zimmermann}},
  \bibinfo{journal}{Nucl. Phys.} \textbf{\bibinfo{volume}{B404}},
  \bibinfo{pages}{517} (\bibinfo{year}{1993}), \eprint{hep-lat/9206025}.

\bibitem[{\citenamefont{Gockeler et~al.}(1991)\citenamefont{Gockeler, Jansen,
  and Neuhaus}}]{Gockeler:1991sj}
\bibinfo{author}{\bibfnamefont{M.}~\bibnamefont{Gockeler}},
  \bibinfo{author}{\bibfnamefont{K.}~\bibnamefont{Jansen}}, \bibnamefont{and}
  \bibinfo{author}{\bibfnamefont{T.}~\bibnamefont{Neuhaus}},
  \bibinfo{journal}{Phys. Lett.} \textbf{\bibinfo{volume}{B273}},
  \bibinfo{pages}{450} (\bibinfo{year}{1991}).

\bibitem[{\citenamefont{Hasenfratz et~al.}(1991)}]{Hasenfratz:1990fu}
\bibinfo{author}{\bibfnamefont{A.}~\bibnamefont{Hasenfratz}}
  \bibnamefont{et~al.}, \bibinfo{journal}{Nucl. Phys.}
  \textbf{\bibinfo{volume}{B356}}, \bibinfo{pages}{332} (\bibinfo{year}{1991}).

\end{thebibliography}

\end{document}